\documentclass[12pt,preprint]{aastex}
\def\lya{Ly-$\alpha$}

\begin{document}

\title{Ly$\alpha$ emitting galaxies at redshift $z$ $\sim$ 4.5 in the LALA Cetus field}

\author{Jun-Xian Wang\altaffilmark{1,2,5}, Sangeeta Malhotra\altaffilmark{2},
James E. Rhoads\altaffilmark{2}, Hao-Tong Zhang\altaffilmark{3}, Steven L. Finkelstein\altaffilmark{4}}
\begin{abstract}
We present a large sample of \lya\ emitting galaxies
spectroscopically confirmed at redshift $z\approx 4.5$,
based on IMACS spectroscopic observations of 
candidate $z$ $\approx$ 4.5 Ly$\alpha$-emitting galaxies in the Large Area Lyman
Alpha (LALA) narrow band imaging survey Cetus field. 
We identify 110 of them as $z\approx 4.5 ~{\rm Ly}\alpha$ emitters based
on single line detections with no continuum emission blueward of the line. Six
foreground galaxies are identified, either based on multiple lines or
blueward continuum emission. The \lya\ confirmation rate varies from $<$50\% to
76\% for candidates selected in different narrowband filters at
slightly different redshifts.  We find a drop in the LAE density at redshift
4.50$\pm$0.03 from redshift 4.39 -- 4.47 by a factor of 66\%, which
could be a large scale void in the distribution of star-forming galaxies 
($\sim 18$ Mpc along the line of sight and $\sim 80$ Mpc across).
The sample includes many objects with equivalent widths (EW) $\ga 200$\AA.
These large EW candidates are spectroscopically confirmed at the same rate
as candidates with more modest EWs. 
A composite spectrum of all 110 confirmed LAEs shows the characteristic
asymmetry of the \lya\ line.  It also places new and stringent upper limits 
on the CIV 1549\AA$/$\lya\ and HeII 1640\AA$/$\lya\ line ratios, 
providing a new upper limit on the fraction of active galactic nuclei
in \lya\ selected galaxy samples, and on the contribution of Pop III populations.  Finally, we calculate the \lya\ luminosity
function for our $z\approx 4.5$ sample, which is consistent with those at other redshifts, showing that there is no evolution in Ly$\alpha$ luminosity function from $z$ = 3.1 -- 6.6.
\end{abstract}


\altaffiltext{1}{Center for Astrophysics, University of Science and Technology of China, Hefei, Anhui 230026, P. R. China; jxw@ustc.edu.cn.}
\altaffiltext{2}{School of Earth and Space Exploration, Arizona State University, Tempe, AZ 85287}
\altaffiltext{2}{National Astronomical Observatories, Chinese Academy of Sciences, 20A Datun Road, 100012 Beijing, China}
\altaffiltext{4}{George P. and Cynthia W. Mitchell Institute for Fundamental Physics  and Astronomy, and Department of Physics, Texas A\&M University, College Station, TX  77843-4242}
\altaffiltext{5}{Key Laboratory for Research in Galaxies and Cosmology, University of Sciences and Technology of China, Chinese Academy of
Sciences}
\section {Introduction}

The number of known high redshift galaxies 
has grown rapidly in the past decade, thanks to efficient wide field
cameras and effective selection techniques (i.e., the dropout technique 
and Ly$\alpha$ line search technique). Narrowband imaging surveys
have become the most efficient approach for finding high redshift \lya\ 
emission lines galaxies.
Great successes have been achieved recently in detecting high redshift 
Ly$\alpha$ emitters using this technique (e.g.
from $z$ = 3 to $z$ = 6.96, 
see Iye et al. 2006;
Rhoads et al. 2000, 2003, 2004; 
Wang, Malhotra \& Rhoads  2005; 
Tapken et al. 2006; 
Westra et al. 2006; 
Shimasaku et al. 2004, 2006; 
Kashikawa et al. 2006; 
Ajiki et al. 2006; 
Ouchi et al. 2005;
Hu et al. 2002, 2004;
Dawson et al. 2004, 2007;
and references therein).
Large well defined samples of high redshift Ly$\alpha$ emitting
galaxies are essential to study the nature of these galaxies (stellar
population, mass etc, e.g. Lai et al. 2008; Pirzkal et al. 2007; Finkelstein et al. 2007; 2009a), the mechanism of Ly$\alpha$ production (Malhotra \& Rhoads 2002) and escape (Finkelstein et al. 2008), the luminosity function and its evolution (Dawson et al. 2007; Malhotra \& Rhoads 2004, 2006; Ouchi et al. 2008), and spatial clustering of star forming
galaxies in the early universe (Wang et al 2005; Ouchi et al 2005; Malhotra et al. 2005; Kova$\check{c}$ et al. 2007).

The Large Area Lyman Alpha (LALA) survey (e.g., Rhoads et al. 2000)
has obtained deep narrowband images in several fields, yielding a large
sample of Ly$\alpha$ emitters at $z$ $\approx$ 4.5 (Dawson et al. 2004; 2007), 
and  $z$ $\approx$ 5.7 (Rhoads \& Malhotra 2001;
Rhoads et al. 2003; Wang et al. 2005), plus a smaller sample
at $z$ $\approx$ 6.5 (Rhoads et al. 2004). 
In this paper, we present the spectra of 110 confirmed Ly$\alpha$ emitters
at $z$ $\approx$ 4.5 in LALA Cetus field with the Inamori-Magellan Areal Camera
and Spectrograph (IMACS, Dressler et al. 2006) on the 6.5 meter Magellan I Baade
Telescope. Together with our previously confirmed sources from Keck
observations (Dawson et al. 2004, 2007), we now have an overall sample
of 171 spectroscopically confirmed Ly$\alpha$ emitters at $z$ $\approx$ 4.5. 

\section{Observations}
\subsection{Candidate Selection through Narrow-Band Imaging}

With the 8192$\times$8192 pixel Mosaic CCD camera on the 4m Mayall Telescope 
at Kitt Peak National Observatory, we have obtained five deep narrow band images 
(36$\arcmin\times36\arcmin$)
in the LALA Cetus field 
(02:05:20 -04:55; J2000.0) to search for candidate Ly$\alpha$-emitters at
$z$ $\approx$ 4.5. The five narrowband filters utilized are overlapping with full
width at half maximum (FWHM) $\approx$ 80\AA\ (see Fig. 1 of Dawson et al.
2004). The central wavelengths are 6559, 6611, 6650, 6692, and 6730\AA\ 
for narrowband filter Ha0, Ha4, Ha8, Ha12 and Ha16 respectively.
The filters thus cover a total redshift range of 4.36 $<$ $z$ $<$ 4.57 for
\lya, with approximately 50\% overlap in bandpass between adjacent filters.
The average seeing in the final stacked narrowband images
are 1.00\arcsec, 1.26\arcsec, 1.12\arcsec, 1.27\arcsec and 0.99\arcsec, and 
the 5 $\sigma$ (2.4\arcsec diameter aperture) limiting magnitudes are 24.7, 24.8, 24.8, 24.7 and 24.1 (Vega) respectively. The zero points of the narrowband images were obtained by matching the narrowband photometry to that of the underlying broadband ($R$) image.

To study the continuum emission properties of our candidates, we
used deep broadband images in $B_W,~R,~I$ band from the NOAO Deep Wide 
Field Survey (NDWFS; Jannuzi \& Dey 1999).  We followed reasonably standard
practices in our narrowband imaging data reduction and our \lya\ galaxy
candidate selection, which are described in detail elsewhere 
(Rhoads et al 2000, Rhoads \& Malhotra 2001, 
Wang, Rhoads \& Malhotra 2005). 
Briefly, our selection criteria are: A) $>$ 5$\sigma$ detection in a 
narrowband filter; B) a narrowband excess of $>$ 0.75 magnitude, so that
$>$ 50\% of the narrowband flux comes from an emission line; C)
significance of the narrowband excess $>$ 4$\sigma$; and D) $<$
2$\sigma$ detection in $B_W$ band. 

We identified a total of 226 candidate Ly$\alpha$-emitters at $z \approx 4.5$
in our primary sample.  
The numbers of candidates selected in each narrow band filter are presented
in Table 1.
Additionally, we performed an extended catalog selection using the spectral
overlap of adjacent filters to go to fainter flux levels.
In this case, we extended the first criterion to $>$ 3$\sigma$ detection in one
narrowband filter but $>$ 5$\sigma$ when stacking two spectrally overlapping
filters. Other criteria are the same. This leads to another 38 candidates, which
comprise our extended catalog.

\subsection{Spectroscopic Observations}

In 2005 September 3--5 we obtained spectroscopy of 194 candidate
$z\approx 4.5$ \lya\ emitters 
in the LALA Cetus field (167 from our primary sample, plus 27 from 
the extended catalog) using the Inamori-Magellan Areal Camera 
and Spectrograph (IMACS, Dressler et al. 2006) on the 6.5 meter 
Magellan I Baade Telescope. 
We used the short camera (f/2, with a 27.2$\arcmin$ diameter
field of view), and used the 300 line grism ($\lambda_{blaze}$ = 6700 
\AA, 1.34 \AA/pixel) in combination with a 5650--9200\AA\ blocking filter
to prevent overlap of higher order spectra.  
All observations employed $10.0\arcsec \times 1.0\arcsec$ slitlets.
The IMACS f/2 pixel scale is $0.2\arcsec$/pixel.  The corresponding 
spectral resolution is $\approx 7$\AA (though it could be somewhat
better for compacat objects observed in subarcsecond seeing).
Five multislit masks (see Table 2) were observed for 2.5 to 3.3 hours in 0.5 hr 
increments. Thanks to the large field of view of IMACS, we were
able to target around 50 candidate Ly$\alpha$-emitters 
in each mask, along with roughly 120 other targets. Of these,
45 candidates were covered by more than one masks. 

The data were reduced using the IMACS version of the 
Carnegie Observatories System for MultiObject Spectroscopy (COSMOS) data 
reduction package\footnote
{http://llama.lco.cl/$\sim$aoemler/COSMOS.html}. 
We first determined 2d wavelength solutions for each science
exposure using arc lamp exposures taken immediately before or after.
We found relatively large wavelength residuals ($\sim 6$ pixels) in the
course of this calibration, perhaps pointing to inaccuracies in the 
geometric coefficients describing the layout of the detector chips
in the focal plane.  Regardless of their origin, these residuals 
are repeatable from exposure to exposure.  They therefore
do not significantly affect our search for Ly$\alpha$ emission lines,
though they do lead to extra uncertainties in the line
wavelengths we determine (see below for details) and sky substraction.

After wavelength calibration, each frame was first bias-subtracted and
flat fielded.  We then performed sky subtraction following the
algorithm described by Kelson (2003), and extracted 1D spectra from
the 2D spectra using the task ``extract-2dspec'' for each slit.
Frequently, this procedure produced an apparently ``negative'' blue
continuum in the 1D spectra of our confirmed Ly$\alpha$ emitters,
indicating small systematic errors in the software's measured sky
level.  In these cases, we checked that the continuum level appears to
be zero in the 2D spectra, and then applied an additive correction to
the 1D spectrum.  We note that none of our subsequent scientific
results rely on the continuum level measured in the spectroscopic data.

To control for possible spatial shifts along the slits between 
individual exposures of a mask, we measured the trace locations of
the 30  brightest continuum sources in the mask.  We corrected for
any measured shifts while stacking the exposures for each mask,
to increase the quality of the stacked 2D spectra.  We also
identified and removed cosmic ray hits while stacking the multiple 
exposures of each mask.
Finally, we extracted the 1D spectra (2\AA\ per bin) using the gaussian extraction
algorithm. 

\section{Spectroscopic Results}
\subsection{Line Identification}

\begin{figure}
\epsscale{1.0}
\plotone{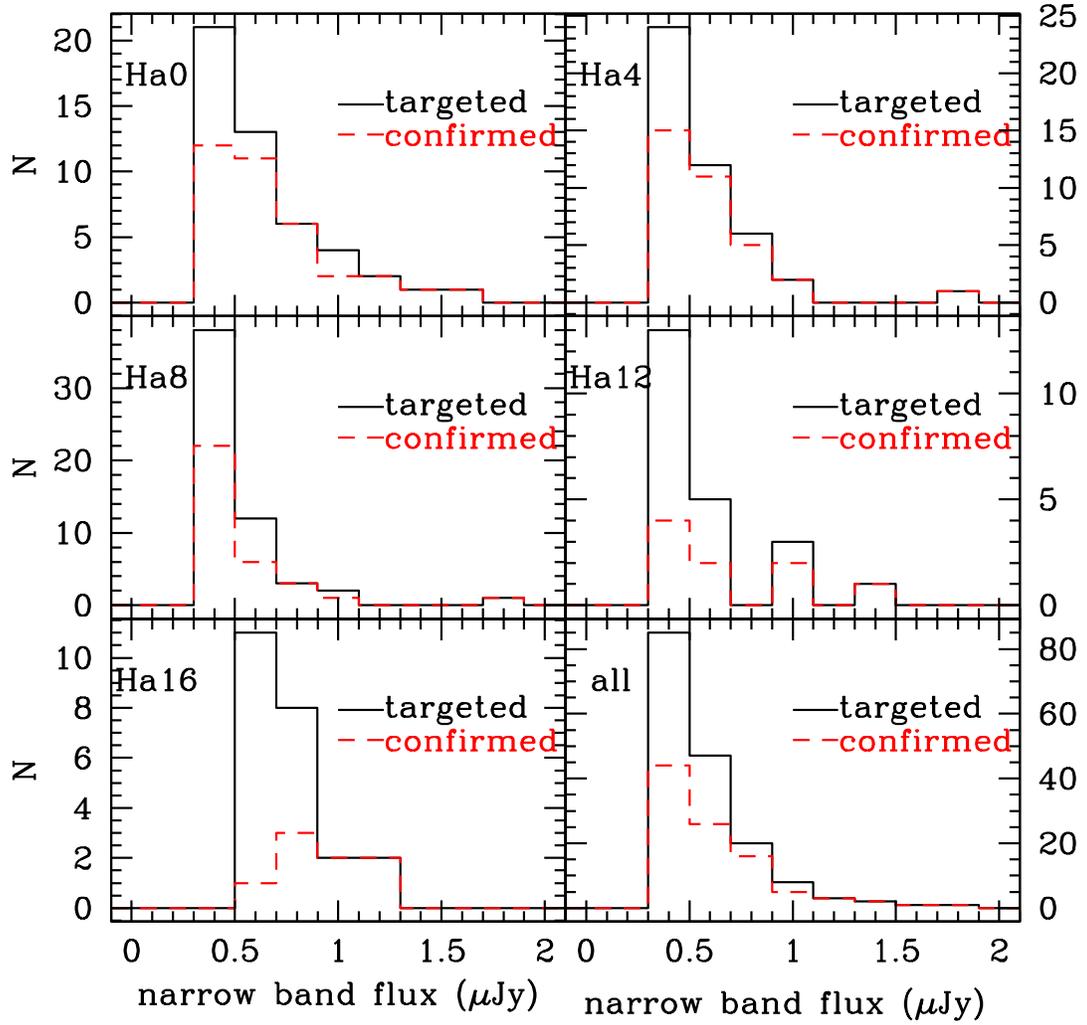}
\caption{The narrowband flux distributions of IMACS targeted candidate (solid) and confirmed Ly$\alpha$ emitters (dashed). Different panels plot the distributions of candidates selected in different narrow bands.
} \label{fluxdis}
\end{figure}

\begin{figure} 
\epsscale{1.0}
\plotone{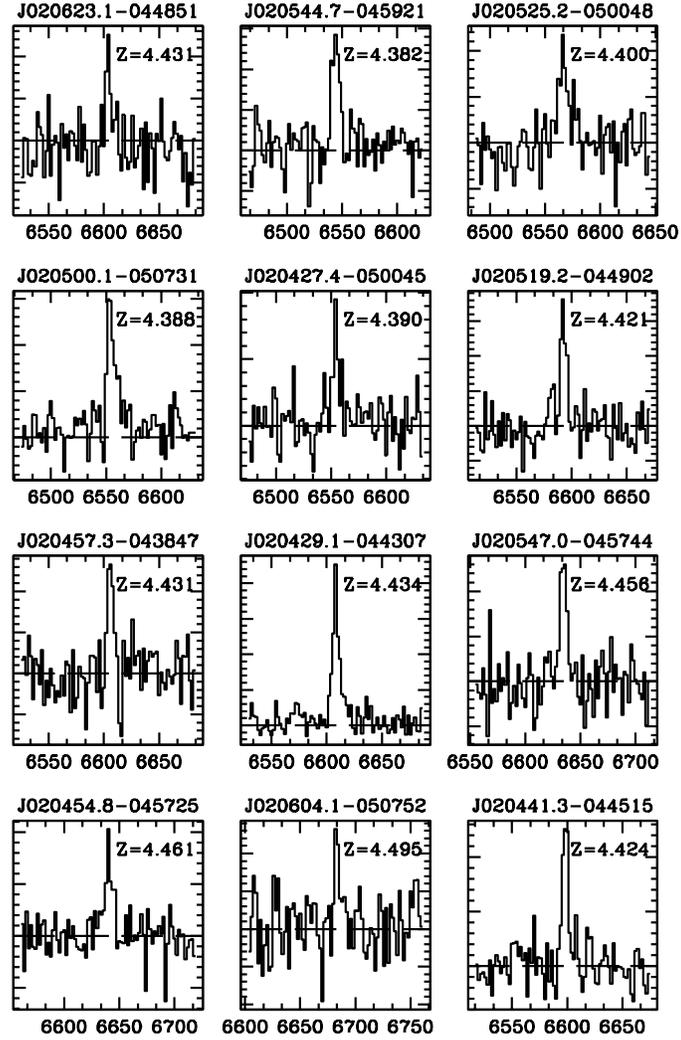}
\caption{Sample IMACS spectra for Ly$\alpha$ line from the set of 110 confirmed LAEs at $z$ $\approx$ 4.5 in LALA Cetus field. 
} \label{spectra}
\end{figure}

We identified 110 single emission lines as Ly$\alpha$ at
$z$ $\approx$ 4.5, among our 194 observed candidates.  Of these, 97
are new spectroscopic confirmations. We present a catalog of the 110
confirmed LAEs at $z$ $\approx$ 4.5 in LALA Cetus field in Table 3.
In Fig. \ref{fluxdis} we presented
the narrowband flux distributions of the targeted and confirmed LAEs
in different narrowbands, and in Fig. \ref{spectra} we present a set
of sample IMACS spectra of confirmed LAEs.

We also identify 6 foreground sources which either show multiple lines
or blue continuum.  The remaining 78 targets were classified as
nondetections.  Of the nondetections, there are 10 whose expected
emission line wavelength range (based on the narrowband data) is
covered or partially covered by the IMACS ccd chip gap (which has 
a width of $\approx$ 80 \AA).  These ten include
one source previously confirmed as a Ly$\alpha$ emitter with Keck LRIS
spectrum by Dawson et al. (2004). The non-detections could

\begin{figure}
\epsscale{1.0}
\plotone{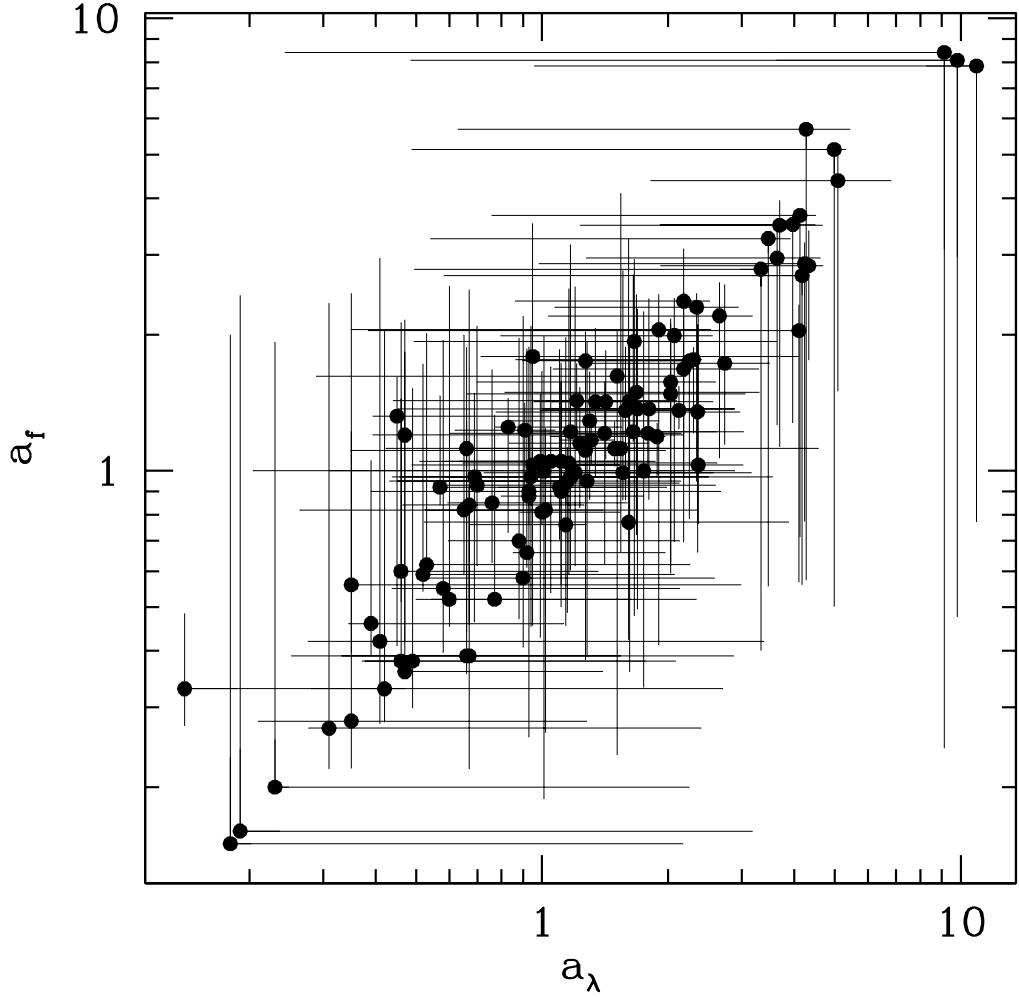}
\caption{The plot of two measurements of the line asymmetry for 110 Ly$\alpha$
emitters (see text for definitions of $a_{\lambda}$ and $a_f$). Due to the low
resolution of our IMACS spectra, both measurements suffer large uncertainties.
Considering the large errorbars, all but 1 source satisfies (within 1$\sigma$)
$a_{\lambda}$ $>$ 1 or $a_f$ $>$ 1, and all but 3 sources satify both.
} \label{asym}
\end{figure}

Following Rhoads et al. (2003, 2004) we perform two measurements of the
line asymmetry. For both measurements, we first determine the wavelength
of the emission line peak ($\lambda_p$), and the data points where the line
flux exceeds 10\% of the peak on the blue side ($\lambda_{10,b}$) and the
red side ($\lambda_{10,r}$). The ``wavelength ratio'' is defined as
$a_\lambda$ = ($\lambda_{10,r}$-$\lambda_p$)/($\lambda_p$-$\lambda_{10,b}$),
and the ``flux ratio'' $a_f$ as the ratio of the accumulated line flux between
$\lambda_{10,r}$ and $\lambda_p$ to that of between $\lambda_p$ and $\lambda_{10,b}$.
In Fig. \ref{asym} we plot $a_\lambda$ versus $a_f$ for the 110 single lines
emitters. The errorbars are obtained through one thousand Monte Carlo 
simulations in which we added random noise to each data bin.
Both measurements are expected to be $>$ 1 for asymmetric profile of high redshift Ly$\alpha$ emission line due to the absorption by neutral hydrogen in outflows from star-forming galaxies and IGM. 

Due to the low resolution and limited spectral quality of our IMACS spectra, 
for most of the lines, we are unable to secure identification of asymmetric profile, i.e., with 
$a_\lambda$ or $a_f$ $>$1 at statistically significant level.
This can be clearly seen from the large errorbars of $a_\lambda$ and $a_f$ in Fig. \ref{asym}.
Considering the large errorbars, we find all but 1 sources statistically satisfy (within 1$\sigma$) $a_{\lambda}$ $>$ 1 or $a_f$ $>$ 1, and all but 3 sources satisfy both. 
This indicates that the spectral quality is insufficient for us to secure or exclude Ly$\alpha$ identification based on line asymmetry measurements.

Possible contaminations to our identifications of Ly$\alpha$ lines are [O II] $\lambda$3727 and
[O III] $\lambda$5007. While [O III] $\lambda$5007 could be typically identified by neighboring [O III] $\lambda$4959 and H$\beta$, [O II] $\lambda$3727 could be identified by looking for corresponding H$\beta$ and [O III] $\lambda$5007 at longer wavelength. These identifications have been done for more than 90\% of the spectra.  For the rest the search for H$\beta$ and [O III] $\lambda$5007 is not possible because of the limited spectral coverage (or due to strong sky emission lines). For these sources, secure identification of [O II] $\lambda$3727 required higher spectral resolution to resolve the doublet, which is not possible by our IMACS spectra.
Luckily, most of the foreground [O II] $\lambda$3727 emission line galaxies have been excluded from our candidate list by excluding sources detected in blueward broadband, and the contamination of [O II] $\lambda$3727 in spectroscopically securely confirmed sample was verified to be very low (e.g. Dawson et al. 2004; 2007). For instance, with Keck DEIMOS spectra, Dawson et al. (2007) have identified two [O II] $\lambda$3727 versus 59 Ly$\alpha$ lines based on line profile, among candidate LAEs mostly selected similarly in LALA Bo\"otes field.
In this paper, we identify all the 110 single lines as Ly$\alpha$ lines.

\subsection{Comparison with Dawson et al.}

\begin{figure} 
\epsscale{1.0}
\plotone{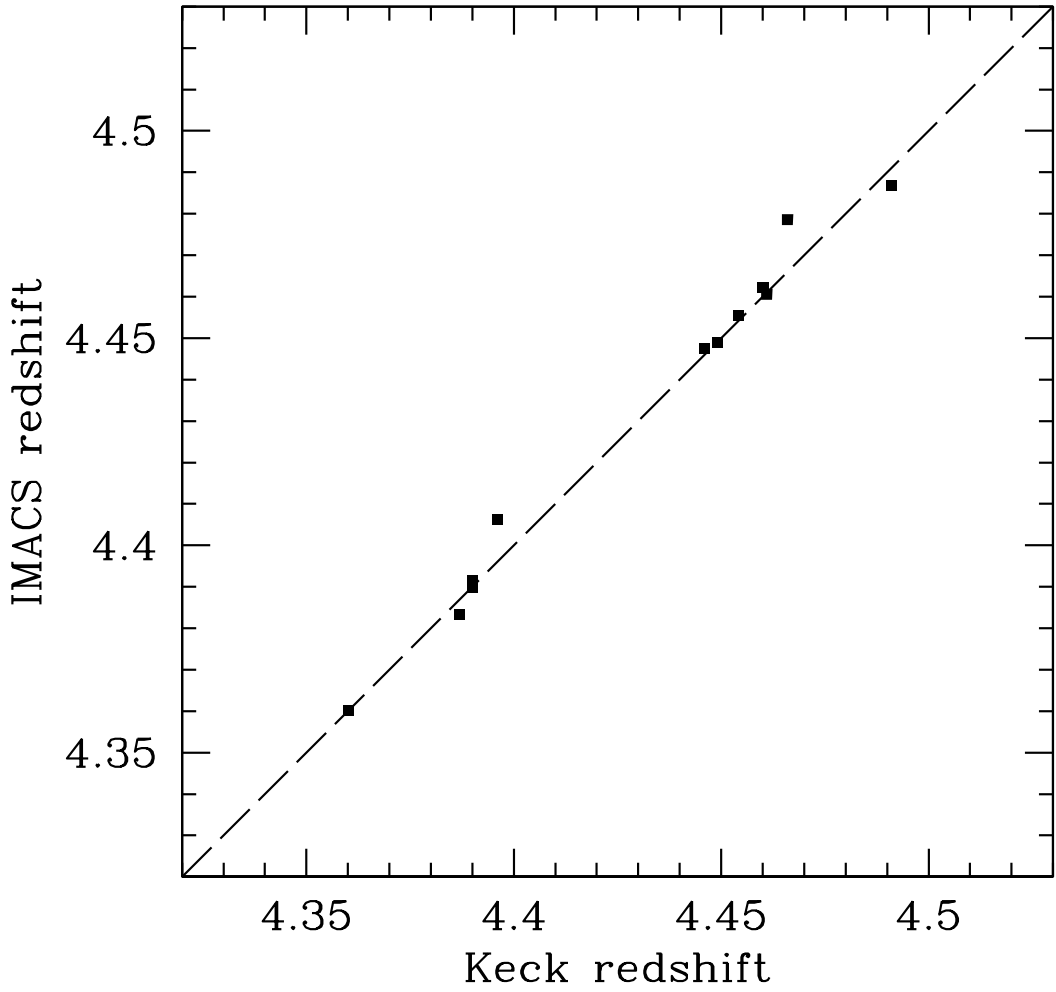}
\caption{The IMACS redshift versus Keck redshift for 13 previously confirmed LAEs with Keck spectra, demonstrating the uncertainty of IMACS wavelength calibration.
} \label{redshift}
\end{figure}

With previous Keck observations, Dawson et al. (2004; 2007) have confirmed a total of 17 LAEs at $z$ $\approx$ 4.5 in LALA Cetus field. While designing our IMACS masks, we gave lower priorities to these previous confirmed LAEs, in order to maximize the total number of LAE candidates with spectroscopic coverage. Thanks to the large field of view of IMACS, we still have 13 of previous confirmed LAEs covered by IMACS observations. 12 of the 13 sources are confirmed at $z$ $\approx$ 4.5 independently with the IMACS observations. The only one without IMACS confirmation has the expected line wavelength covered by IMACS chip gap. In Fig. \ref{redshift} we plot the observed Keck 
redshifts versus IMACS redshifts for these 12 sources, which demonstrates the uncertainties in the IMACS wavelength calibration ($\delta$$z_{1\sigma}$ = 0.005 at $z$ $\approx$ 4.5).
Our IMACS exposures also covered 7 candidates in Cetus field which were previously reported as non-detections with Keck spectra (Dawson et al. 2004; 2007). 
None of these candidates showed any emission line in the IMACS spectra either. These results suggest that our IMACS spectra achieve sensitivities comparable
to our previous Keck observations.   
The 17 confirmations reported by Dawson et al (2004; 2007) were drawn
from a total of 26 LAE candidates observed with Keck in the LALA Cetus
field, for a success rate of 65\%. We note that 24 out of these 26 candidates
were selected from Ha0, Ha4 or Ha8 band (since final, full-depth reductions
of the Ha12 and Ha16 filter images were not available at the time of
the Keck observations).   For the Ha0, Ha4 and Ha8 bands alone, the
success rate of Dawson et al. is 71\%.   Our IMACS success
rate for these same three filters is 66\%, consistent with Dawson et al.

\subsection{Equivalent width distribution}
In our previous studies among others(e.g. Malhotra \& Rhoads 2002; Dawson et al. 2004; 2007; Kudritzki et al. 2000), large fractions of candidate LAEs were reported to show rest-frame Ly$\alpha$ line EW above 240 \AA, which is too large for normal stellar populations, unless they have a top-heavy initial mass function (IMF), zero (or very low) metallicity, and/or extreme youth (age $<$ 10$^6$ yr). Significant contributions of type 2 active galactic nuclei (AGNs) to the large line EW has been ruled out (see \S4.2 below). The large line EW could also likely be due to the clumpy and dusty inter-stellar medium in the galaxy which could attenuate the continuum photons but resonantly scatter the line photons at the surface of the dusty clouds, thus increase the observed EW (Neufeld 1991; Hansen \& Oh 2006; Finkelstein et al. 2007; 2008; 2009).

\begin{figure} 
\epsscale{1.0}
\plotone{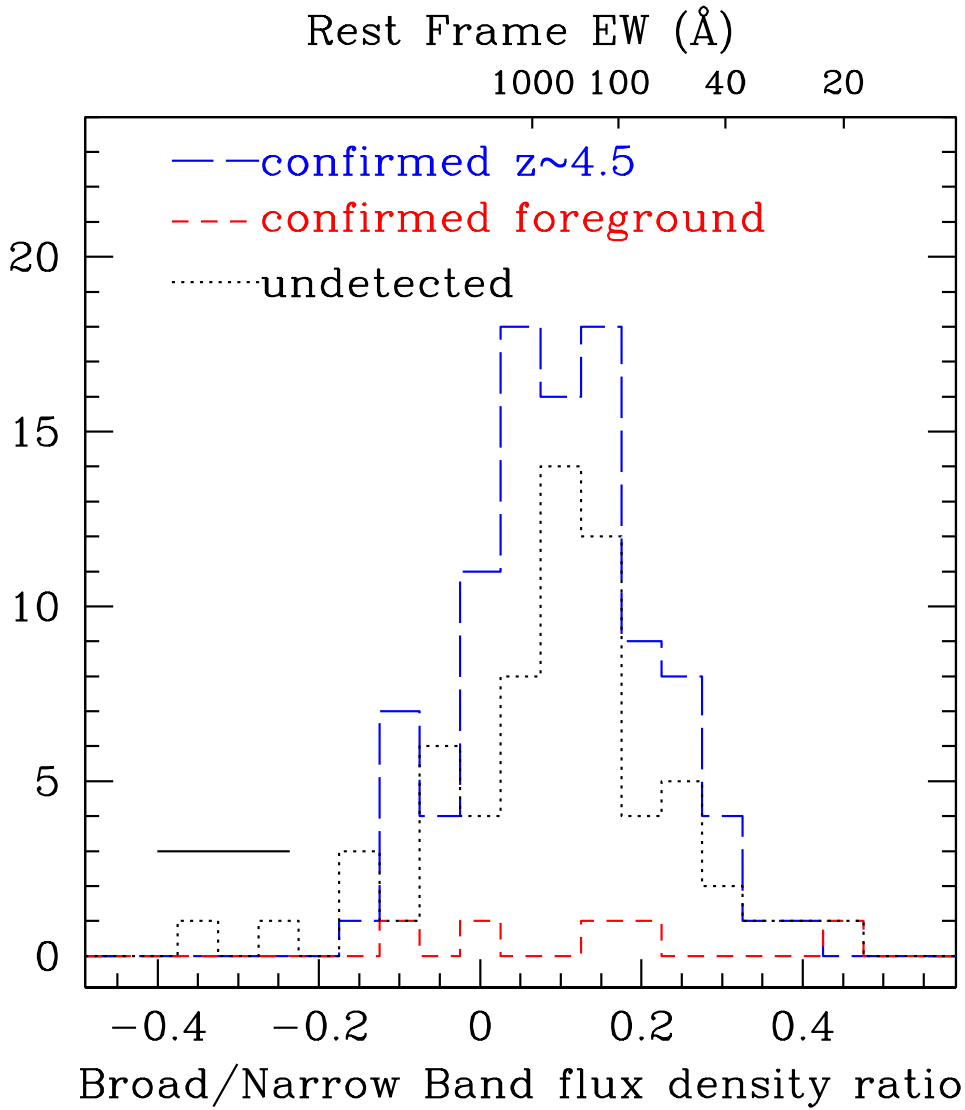}
\caption{The distribution of the broad to narrow band flux density ratio for the spectroscopically confirmed z $\approx$ 4.5 sources, foreground sources, and undetected sources.
The typical error in the flux density ratio is plotted in the bottom left corner.
The line rest frame line EW (marked on the top of the plot) is a monotonic decreasing function of the flux density ratio. From the figure we see no difference in the flux density ratio distribution of pectroscopically confirmed z $\approx$ 4.5 sources and undetected sources. 
The KS test yields no difference in these two populations ($p$ = 85\%).
This suggests the EW distribution of all candidates or all confirmed z $\approx$ 4.5 sources can represent that of all true z $\approx$ 4.5 sources in our sample.
} 
\label{ewhist}
\end{figure}

Could the candidate LAEs with higher EW based on photometry (i.e. these with no continuum detection) be contaminations due to noise spikes, or spurious features? 
Our large spectroscopic sample enable us to perform a statistical test on this issue.
In Fig. \ref{ewhist} we plot the distributions of the broad to narrow band flux density ratio for our confirmed $z$ $\approx$ 4.5 LAEs, foreground sources and spectroscopically undetected sources. The Ly$\alpha$ line EW is a monotonic decreasing function of the broad to narrow band flux density ratio. In the figure we can see that these spectroscopically confirmed LAEs and undetected sources have a consistent distributions of broad to narrow band flux density ratio. 
A Kolmogorov-Smirnov test shows that two distributions are indistinguishable ($p$ = 85\%).
This demonstrates that high EW candidates are confirmed at the same rate.

In table 3 we also present the line equivalent width for all the spectroscopically confirmed z $\approx$ 4.5 galaxies. Due to the large uncertainty in estimating the continuum level from optical spectra, we utilize narrow and broad band photometry to derive the line equivalent width. 
Following Maholtra \& Rhoads (2002), we calculate the rest frame line EW as F$_{Ly\alpha}$/$f_\lambda$/(1+$z$)=($W_RN$ - $W_NR$)/($R$ - $N$)/(1+$z$), where F$_{Ly\alpha}$ is the Ly$\alpha$ line flux, $f_\lambda$ the continuum flux density, $W_N$ and $W_R$ the widths of the $R$ band (1568\AA) and the narrow band (80\AA), and $R$ and $N$ the integrated fluxes in the $R$ band and narrow band respectively. 

Consistent with Malhotra \& Rhoads (2002), we find only 50\% of the confirmed LAEs show measured line EW $<$ 240\AA\ in the rest frame. The rest either show larger EW or no detection of continuum at all. It's clear that the uncertainty in line EW mainly comes from that of the continuum, especially for high EW sources, we thus calculate the uncertainty of the line EW by applying 1$\sigma$ errorbar to the measured $R$ band flux. In many cases that the upper limits of the line EW can not be constrained, lower limits are given.

We point out that the line EW distributions of different selected samples are sensitive to the selection criteria adopted. Particularly,  the requirement of broad band detection (i.e. Hu et al. 2004) would exclude very high EW sources. 
For instance, only 1 out of 7 LAEs at redshift 3.7 selected by Fujita et al. (2003) in Subaru/XMM-Newton Deep Field show rest frame line EW above 200\AA\ (based on narrow and broad band photometry), however, this is after they excluded the candidates without broadband detection, most of which (6 out 7), if located at
redshift 3.7, show line EW above 200\AA, consistent with LALA surveys.
The EW distribution is 
also likely sensitive to the relative depths of the broad and narrow band images, e.g., deeper broad band images could enable detection of more sources with smaller EWs because more sources with stronger continuum could pass the selection due to smaller broadband flux uncertainty. 

We have presented above that most of the sources with high EWs have no broadband detection at all.
It's interesting to note that if we sum the broad and narrow band fluxes for all the confirmed LAEs, we obtained an average EW of 182$^{+26}_{-20}$\AA. If we sum the broad and narrow band fluxes for the confirmed LAEs with $<$ 3$\sigma$ continuum detection (100 out of 110), we derived an average EW of 316$^{+102}_{-62}$\AA. This also demonstrates that a large population of these individual LAEs have high EW.

\begin{figure} 
\epsscale{1.0}
\plotone{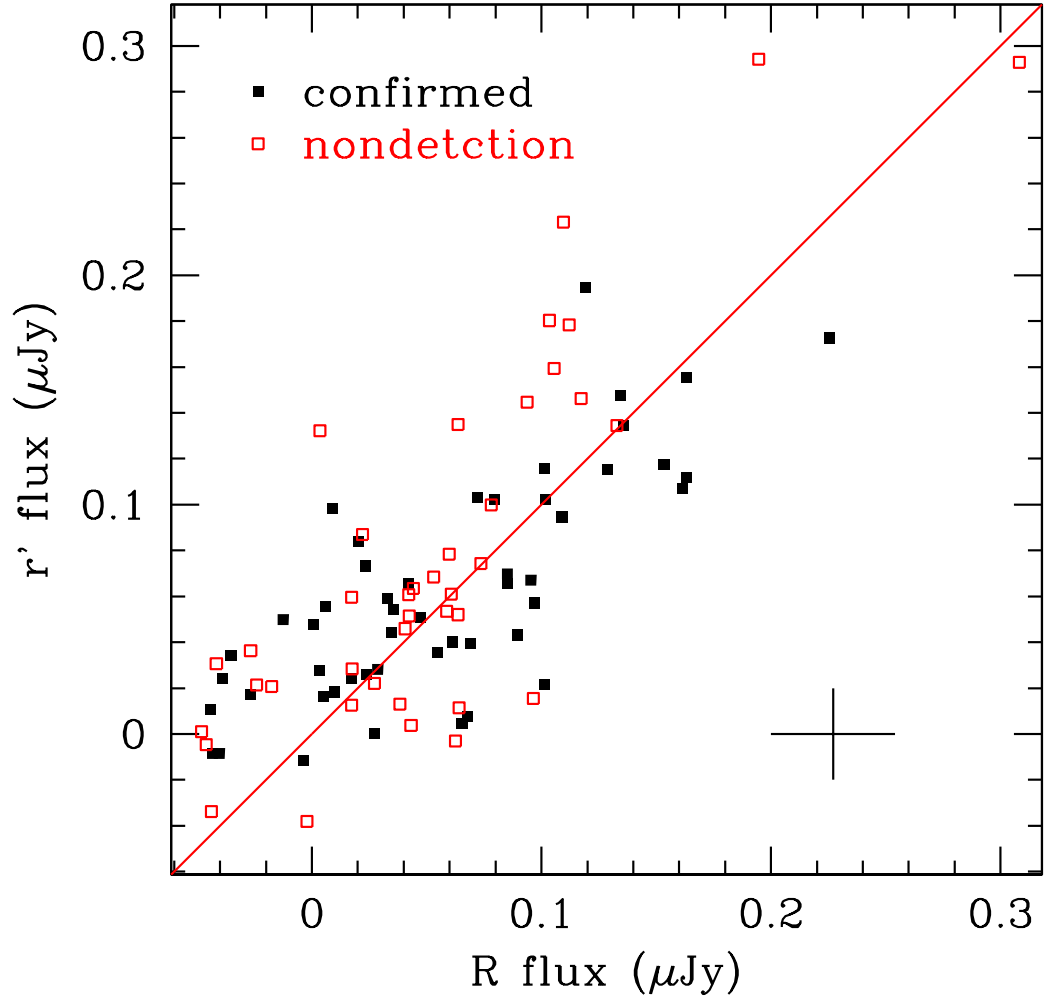}
\caption{The flux density in the MMT r' band versus CTIO R band for 44 confirmed LAEs.
Spectroscopical non-detections are also overplotted.
The solid line represents equal lines in both bands. The typical errors in the measurements are
presented in the bottom right corner.} 
\label{com}
\end{figure}

One final concern on the weak continuum level is whether they are
physically weak or due to selection bias.  That is, did the selection
favor the sources whose broad band fluxes were underestimated due to
random fluctuations? If this were an important effect, there should be
one immediate consequence: these high EW sources would be
systematically brighter in an independent broadband image. We adopted
the independent broadband images obtained with MMT (Finkelstein et
al. 2007; Wang et al. 2007) to test this possibility. In
Fig. \ref{com} we plot CTIO R band flux versus MMT r' ``flux'' for 44
confirmed LAEs with r' band coverage.  Here the MMT r' ``flux'' was
scaled by a factor which was derived by requiring that the clipped
average $R-r'$ color be zero for the full list of sources well
detected in both images.  We note that the band pass of MMT r' band is
not identical to that of CTIO band, however, considering MMT r' band
has much better efficiency between 6500\AA\ to 7200\AA\ than CTIO R
band, while CTIO R band extends to 8000\AA\ with transmission below
40\%, the fluxes from two bands are comparable for our purpose.  From
the figure we can see that these LAEs have consistent r' band fluxes,
suggesting the possible selection bias we mentioned above, if there is
any, is weak.  The clearest exception is that fluxes that appear
negative in the R band image (and whose R fluxes are therefore surely
underestimated due to photon noise) mostly have r' fluxes that
are zero or slightly positive.
EW calculations show that 24 out of 44 confirmed
LAEs with r' band flux measurements have EW above 240\AA\ based on old R band
measurements, while 21 of the 44 show EW above 240\AA\ based on r' band
fluxes. The EW calculations based on r' band fluxes have been adjusted to
the r' band transmission curve.  We conclude that the high EWs we have
obtained are physically real, and cannot be fully ascribed to
selection bias.

\subsection{The stacked spetrum}

\begin{figure} 
\epsscale{1.0}
\plotone{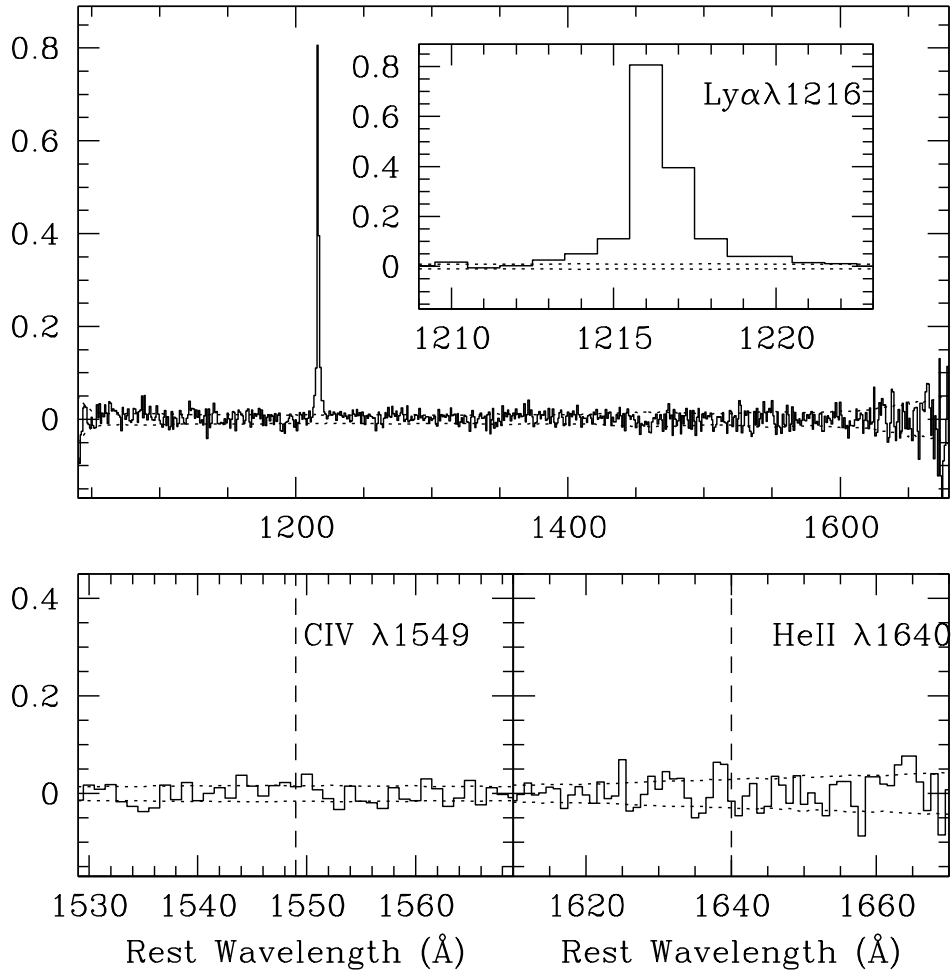}
\caption{The co-added spectrum of the 110 IMACS confirmed Ly$\alpha$ emitters at $z$ $\approx$ 4.5 in the LALA Cetus field.  Dashed lines show the one sigma uncertainty in the composite spectrum. In the co-added spectrum, neither CIV $\lambda$1549 nor He II $\lambda$1640 line could be detected in the vicinity of expected wavelengths, and 2$\sigma$ detection limits were given in the text.
} \label{coadd}
\end{figure}

Stacking a large number of spectra could further help us understand the nature of the line emitters,
especially for those with low signal to noise ratio.
In Fig. \ref{coadd} we present the stacked spectrum of all 110 
confirmed Ly$\alpha$ emitters at $z$ $\approx$ 4.5. We perform a 
variance-weighted co-addition of the 110 spectra. In each wavelength bin,
data points from each spectrum (normalized by the peak flux of the
Ly$\alpha$ 1216\AA\ emission line) were averaged using two sigma
clipping algorithm (one iteration) to remove artifical features due to
sky line residuals, ccd edges, etc. 

In the composite spectrum, the only visible line feature is the asymmetric 
Ly$\alpha$ line (highlighted in the cutout of Fig. \ref{coadd}). 
In the co-added spectrum, we obtained $a_\lambda$ = 1.46$\pm$0.16, 
and $a_f$ = 1.42$\pm$0.09, consistent with those expected from the asymmetric profile
of high redshift Ly$\alpha$ emission line.
The errorbars on $a_\lambda$ and $a_f$ are
also estimated based on large number of Monte Carlo simulations in which we added 
random noise to each data bin.

In the co-added spectrum,  we searched for possible CIV $\lambda$1549 and He II $\lambda$1640 line in the vicinity of expected wavelengths, however, neither of them could be detected at above 2$\sigma$ level.
Based on the flux uncertainties in the spectra, we obtained 2$\sigma$ detection limit (by fixing the line width to that of Ly$\alpha$) to the flux ratio of $f$(CIV $\lambda$1549)/$f$(Ly$\alpha$) and $f$(He II $\lambda$1640)/$f$(Ly$\alpha$), which are 3.7\% and 7.4\% respectively.
These constraints are significantly stronger than we previously obtained based on 11 LAEs spectra with Keck (8\% and 13\% respectively, Dawson et al. 2004). 

\section{Discussion}

\subsection{Large scale structures}

We 
find significantly different success rates and total numbers of candidates obtained in images from different narrowband filters   (see Table 1). The small number of candidates in the Ha16 filter can be ascribed to the shallower depth in the Ha16 narrow band image, and the lower success rate could thus be due to a higher percentage of contamination from noise spikes (assuming the total number of spurious detections due to noise spikes per image remains constant) and asteroids (with a small number of frames, asteroids could leave more spurious features in the stacked image). 

The Ha12 narrowband image has the same depth as Ha0, Ha4 and Ha8 bands, yet has many fewer candidates detected. This indicates that the source density at the redshift 4.50$\pm0.03$ is much smaller than that in the redshift range of 4.39 - 4.47 (covered by Ha0, Ha4 and Ha8 filter). 
The success rates and number of candidates in Ha0, Ha4 and Ha8 filters are similar, with Ha8 having marginally more candidates, and a marginally lower confirmation rate. 

\begin{figure} 
\epsscale{1.0}
\plotone{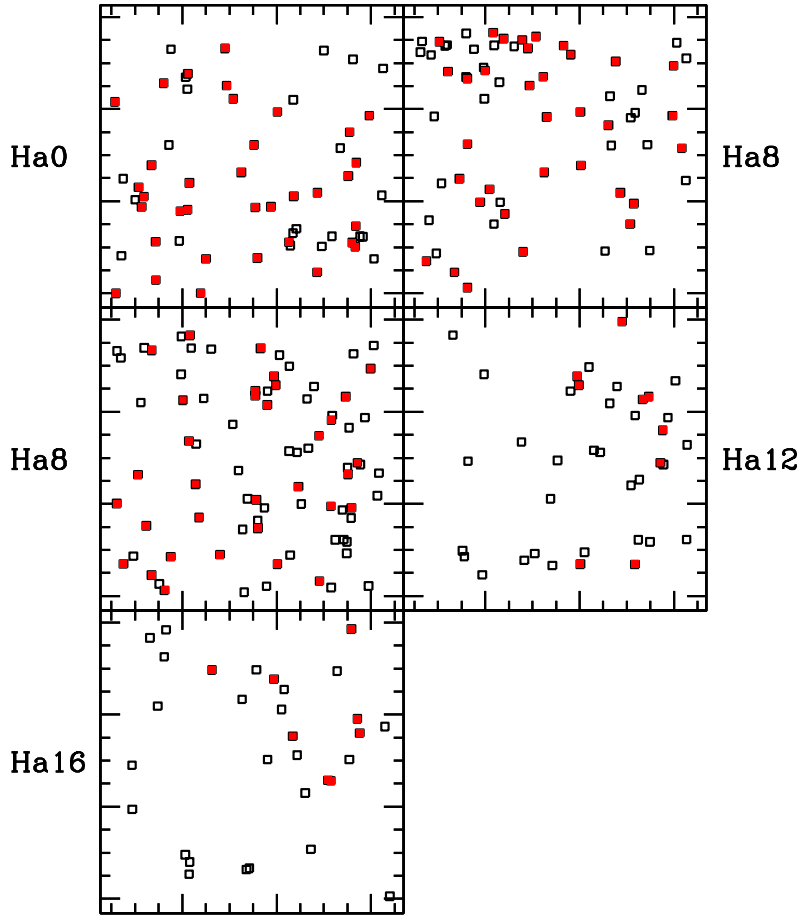}
\caption{The sky distribution of the $z$ $\approx$ 4.5 candidates (confirmed LAEs are over plotted as solid squares) detected in different narrowband. The size of the box is 38$\arcmin$ $\times$ 38$\arcmin$.
The distribution of the confirmed objects projected on the sky is subject to
complex constraints on slit placement in multislit spectroscopy,
so angular clustering in this figure should not be interpreted as
a pure signature of galaxy clustering.
Note except for Ha16 band, which is 0.6-0.7 magnitude shallower, all other four narrow band images have comparable limiting magnitudes. This figure clearly show that the source density of LAEs detected remains constant from redshift 4.39 to 4.47, but shows a significant drop at redshift 4.50. 
} \label{sky}
\end{figure}

In Fig. \ref{sky} we plot the spatial distributions of the candidates selected in each narrowband image.  Similar large scale structure has been reported at various redshifts elsewhere (e.g. see Ouchi et al. 2005, Wang et al. 2005,  
Palunas et al. 2004, Kovac et al. 2007, and more). Quantitative analysis on the large scale structures presented in Fig. \ref{sky} requires careful examination on the candidate selection effect, the coverage of mask design, and the success rate of each mask in each narrow band, etc, which is beyond the scope of this paper. However we note that the amplitude of the drop in LAE source density at redshift 4.50$\pm0.03$ (66\%)\footnote{We first calculated the expected numbers of LAEs in Ha0, Ha4, Ha8 and Ha12 band respectively, by multiplying the total number of candidates with the spectroscopical success rate in each band. The fraction 66\% was obtained  by comparing the expected number of LAEs in Ha12 band with the average value in Ha0, Ha4 and Ha8 band.} is consistent with the cosmic variance calculation (Trenti \& Stiavelli 2008), which gives a 40\% cosmic variance (1$\sigma$) for the volume of our survey.

\subsection{Active Galactic Nucleus(AGN)  Fraction}

Significant contributions of AGNs to the large line EW has been ruled out with deep X-ray exposure and optical spectra (e.g. see Malhotra et al. 2003; Wang et al. 2004; Dawson et al. 2004; Gawiser et al. 2006).  Deep X-ray images can be used to test whether these LAEs are AGNs, since they are expected to be strong X-ray emitters based on their Ly$\alpha$ emission line. However, with two deep Chandra ACIS exposures ($\sim$ 180 ks each), we were unable to detect any of the  101 LAE candidates in LALA Bo\"{o}tes and Cetus field. The stacking technique helps to establish a strong constraint (3$\sigma$) to the average X-ray luminosity of L$_{2-8keV}$ $<$ 2.8 $\times$ 10$^{42}$ ergs s$^{-1}$, and we estimated that $<$ 4.8\% of the LAEs in LALA Bo\"{o}tes and Cetus field could be possible AGNs (Wang et al. 2004).
No AGN lines have been identified in the obtained individual and composite spectra either (Dawson et al. 2004; and this paper).  Dawson et al. (2004) have presented an upper limit of
CIV to Ly$\alpha$ line flux ratio of 8\% (2$\sigma$) based on the composite spectrum of 11 LAEs at $z$ $\approx$ 4.5.

In \S3.4 we have presented the composite spectra of 110 Ly$\alpha$ emitters at $z$ $\approx$ 4.5.
Even in the composite spectrum, CIV line is invisible. A much stronger  upper limit to the CIV line flux is derived, which is $<$ 3.7\% (2$\sigma$) of the Ly$\alpha$ flux, while typical ratio of CIV to Ly$\alpha$ line in type II AGNs\footnote{Type 1 AGN could be ruled out in most (if not all) LAEs because of the lack of broad emission lines in the spectra.} is 22\% (Ferland \& Osterbrock 1986). This confirms previous results that the AGN contribution to LAEs is rather small.   The formal upper limit from the CIV line
is $\la 3.7/22 =17\%$.  This is somewhat weaker than our earlier X-ray
based limits (Malhotra et al 2003, Wang et al 2004), but is an entirely
independent constraint.  
Interestingly, this limit is below the AGN detection fraction for
nearby (redshift $z\sim 0.3$) LAEs, reinforcing the observation of AGN
fraction evolution by Finkelstein et al (2009b).

\subsection{Population III stars}

The search for the first generation (or Pop III stars) is one of the major targets in observational studies. However,  until now there is no strong evidence of the detection of Pop III stars, and the search for such objects remains a major observational challenge (Schaerer 2008). 
Recent numerical calculations have shown that since Lyman continuum flux increases with decreasing metallicity in stellar populations, very low metallicity stellar population can produce strong Ly$\alpha$ line with maximum EW $>$ 400 -- 850\AA\ for $Z$ $<$  10$^{-5}$ $Z\sun$ (e.g. Schaerer 2003; Tumlinson, Shull \& Venkatesan 2003). Besides strong Ly$\alpha$ emission line, Pop III or very low metallicity stellar population is also expected to produce strong He II 
$\lambda$1640 due to its much stronger He II ionizing continuum. Thus many studies have been performed to search for strong He II $\lambda$1640 emission line in the spectra of high redshift LAEs, but no positive detection has been obtained yet.

Consistent with Malhotra \& Rhoads (2002) and Ouchi et al. (2008), we have found a large population of spectroscopically confirmed LAEs with rest frame Ly$\alpha$ line EW above 240\AA. 
We have stated above that these high EWs are real.
This is in contrast to low EWs found at $z$ = 3.1 \& 2.2 (Gronwall et al. 2007; Nilsson et al. 2009), whether this is due to cosmic evolution or different selection criteria is not clear.
We have also ruled out significant contribution of AGN to the large EW (\S4.2). 
Another possibility to the observed large Ly$\alpha$ EW is due to very  metal-poor or Pop III stars.
However,  in both the individual and the co-added IMACS spectra, we detect no evidence 
of He II $\lambda$1640 emission line. In the co-added spectrum, the He II $\lambda$1640 emission line flux is constrained to be $\lesssim$ 7.4\% (2$\sigma$) of that of Ly$\alpha$.
Such an upper limit can be converted to log (Q$_{He+}$/Q$_H$) $<$ - 1.4, where Q$_{He+}$ and Q$_H$ are ionizing fluxes for He+ and H respectively (Schaerer 2003).  This upper limit is close to the maximum predicted values for metal-free populations with IMFs including very massive stars (up to 500 M$\sun$, see Fig. 5 of Schaerer 2003). This indicates that although He II $\lambda$1640 is not detected in our composite spectrum, the derived upper limit  is insufficient yet to rule out the existence  of Pop III  (or very metal-poor) stars in our LAEs at $z$ $\approx$ 4.5.
Considering the fraction of Pop III galaxies among LAE sample at $z$ $\sim$ 4.5 could be rather small (such as a few percent, e.g. Scannapieco, Schneider \& Ferrara 2003),  the derived upper limit can not even rule out some model which predict comparable He II $\lambda$1640 and Ly$\alpha$ line fluxes for Pop III populations (e.g. Tumlinson et al. 2003).

\subsection{Ly$\alpha$ Luminosity Function}

We have presented a large spectroscopically confirmed LAE sample at $z$ $\approx$ 4.5. In this section we present the Ly$\alpha$ luminosity function based on this sample.
Due to the large uncertainty in flux calibration in spectroscopic observations, we adopt photometric measurements to calculate the Ly$\alpha$ fluxes. 
The Ly$\alpha$ fluxes were derived from narrowband and broadband photometry, where $f_{Ly\alpha}$ = $f_{nb}$ - $f_{continuum}$*$w_{nb}$ and $f_{continuum}$ = ($f_R$-$f_{nb}$)/($w_R$-$w_{nb}$). The line luminosities of the confirmed LAEs were calculated assuming a cosmology model of H0 = 71 km/s/Mpc,  $\Omega_m$ = 0.27 and $\Omega_\lambda$ = 0.73.

\begin{figure} 
\epsscale{1.0}
\plotone{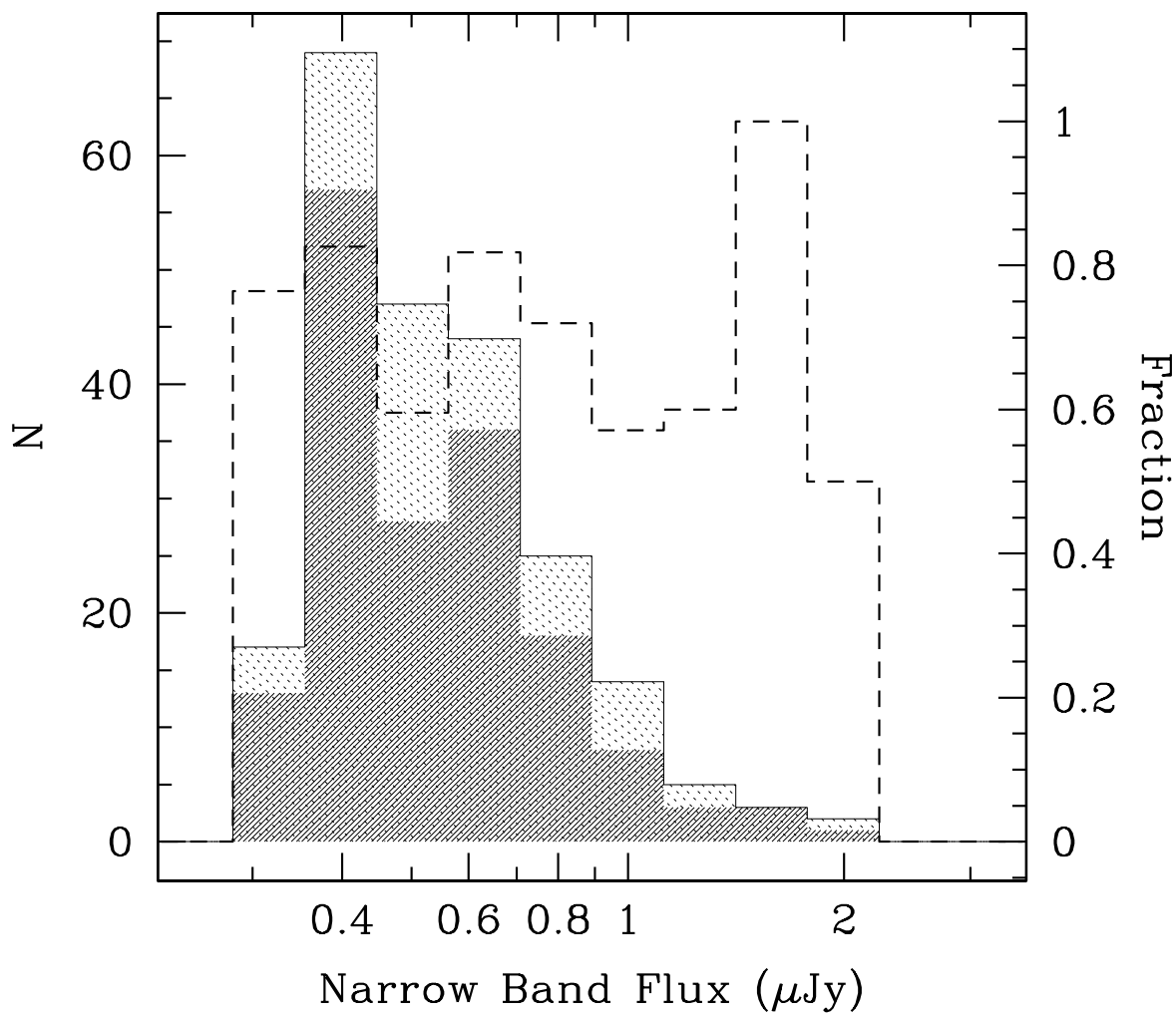}
\caption{The distribution of the narrow band fluxes for all candidate LAEs (solid line), and those targeted for spectroscopy (thick shaded region).
For our narrowband filters, a flux of 1 microJansky corresponds to a line flux of 5.5$\times$10$^{-17}$ erg/cm$^2$/s for an object with no continuum.
The fraction of candidates LAEs with spectroscopic follow-up is over-plotted (dashed line). 
} 
\label{fraction}
\end{figure}

For each confirmed LAE, we first calculated the comoving volume $V_{max}$ over which the source could be selected by our survey from $z_{min}$ to $z_{max}$, which are 4.36 and 4.65 respectively through band Ha0 to band Ha16. The survey area is 35\arcmin\ $\times$ 35\arcmin.
Effects of different flux limits in narrow bands were also taken into account.
We show the incompleteness of our spectroscopical observations in Fig. \ref{fraction}. In the figure we plot the fraction of spectroscopically targeted  LAEs as a function of Ly$\alpha$ line flux. Such incompleteness was taken into account while calculating the luminosity function.
To be conservative, we assume that all the spectroscopically non-detected sources are not LAEs. 

\begin{figure} 
\epsscale{1.0}
\plotone{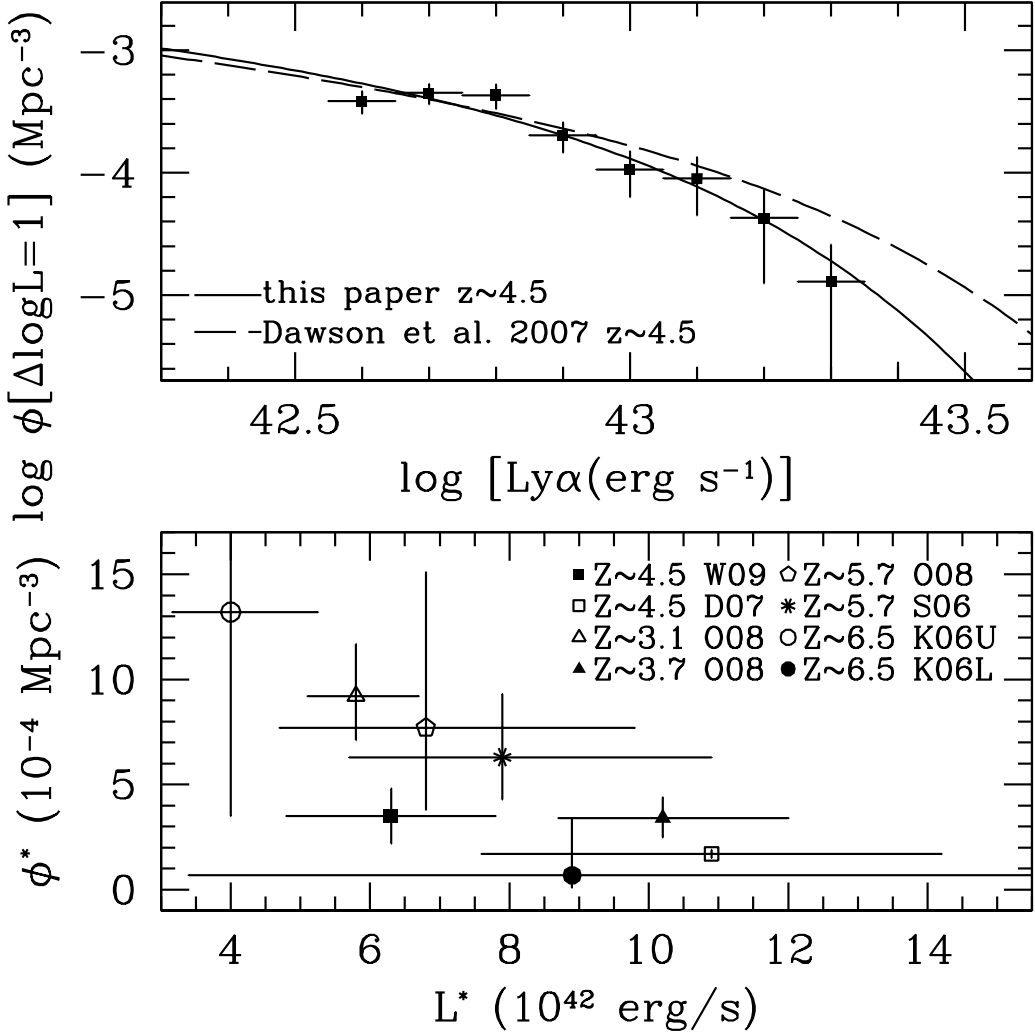}
\caption{The differential luminosity function of LAEs at $z$ $\approx$ 4.5. The best-fit Schechter function with slope fixed at -1.5 is over plotted (solid line). The best fit LF published in Dawson et al. (2007) is also over-plotted for comparison (dashed line). In the lower panel, we compare $L^*$ and $\Phi^*$ from various surveys at different redshifts. W09: this work; D07: Dawson et al. (2007); O08: Ouchi et al. (2008); S06: Shamasaku et al. (2006); K06U: Kashikawa et al. (2006) upper limit; K06L: Kashikawa et al. (2006) lower limit. Best-fitting parameters obtained by fixing $\alpha$ = -1.5 are taken for all surveys except for D06 ($\alpha$ = -1.6).}
\label{lf}
\end{figure}

The derived Ly$\alpha$ luminosity function of LAEs at $z$ $\approx$ 4.5 is plotted in Fig. \ref{lf}.  
Following Malhotra \& Rhoads (2004), we fit the luminosity function with a Schechter function
\begin{equation}
\Phi(L) \, dL = \frac{\Phi^*}{L^*} \left (
\frac{L}{L^*} \right )^\alpha \exp \left ( -
\frac{L}{L^*} \right ) \, dL \; .
\end{equation}
During the fit, we fixed the slope $\alpha$ at -1.5 since our data is insufficient to have it constrained. The best-fit parameters are 
$L^* = 6.3\pm1.5 \times 10^{42}$ erg s$^{-1}$ and $\Phi^* = 3.5\pm1.3 \times 10^{-4}$ Mpc$^{-3}$.
Considering the large uncertainties in the measurements and differences in Ly$\alpha$ emitting galaxies selection criteria among different research groups, the best-fitted $L^* $ and $\Phi^*$
are consistent with the range of values found in various surveys at $3\la z\la 6.5$
(see the lower panel in Fig.  \ref{lf}).

\section{Summary}
We have presented the largest spectroscopic sample of $z\approx 4.5$
\lya\ emitting galaxies yet studied.  Examining the properties of this
sample, we reach several conclusions:\\

(1) Large equivalent width \lya\ emitters, which we have earlier
reported as a common feature of narrowband-selected samples (Malhotra
\& Rhoads 2002), are spectroscopically confirmed at the same rate as
less spectacular line emitters.  Thus, the high equivalent width
sample is not dominated by noise spikes.  Nor can the large equivalent
widths be solely a byproduct of calculating the EW using the same
broad band data employed in candidate selection, because our large EW
subsample changes little when we calculate EW using an independent broad band
image completely independent of object selection.

(2) The CIV 1549\AA\ line is (on average) $\la 3.7\%$ of the \lya\ 
line strength, placing an upper limit of $\la 17\%$ on the AGN fraction
in the \lya\ selected sample.  This limit is independent of the X-ray
to \lya\ ratio in AGN and therefore complements earlier limits on the
AGN fraction based on X-ray photometry.

(3) The HeII 1640\AA\ line is $\la 7.4\%$ of the \lya\ line strength.
This is the most stringent limit to date on the HeII emission from these
objects.  Though it does not strongly rule out many Population III
models yet, it is nearly deep enough to do so and could be combined 
with a few similarly sized samples in future.

(4) We present a new \lya\ luminosity function, based on a large spectroscopic
data set at $z\approx 4.5$.  Allowing for differences
in \lya\ galaxy selection criteria among different research groups,
this luminosity function remains broadly consistent with others over
the range $3\la z\la 6.5$.  

\acknowledgements
The work of JXW is supported by National Basic Research Program of China 
(973 program, Grant No. 2007CB815404),  Knowledge Innovation Program of CAS (Grant No. KJCX2-YW-T05),  and Chinese National Science Foundation (Grant No. 10825312).
Our data were 
obtained using community access telescope time made available
under the National Science Foundation's Telescope System 
Instrumentation Program (TSIP), awarded by the National Optical Astronomy 
Observatory.
The work of JER and SM is supported in part by NSF grant AST-0808165.

\clearpage
\begin{deluxetable}{llllll}
\tabletypesize{\scriptsize}
\tablecaption{Numbers of candidate, targeted and confirmed Ly$\alpha$-emitters
at $z \approx 4.5$ in each narrow band filter in the LALA Cetus field.
The ``ext'' line represents candidates in the extended sample, which goes
to a fainter flux level by combining data from filters that overlap in 
wavelength.\label{}}
\tablewidth{0pt}
\tablehead{
\colhead{filter} &\colhead{average seeing} &\colhead{limiting mag$^a$} &\colhead{candidates} & \colhead{targeted} & \colhead{confirmed}
}
\startdata
Ha0 & 1.00\arcsec& 24.7&57 & 48 & 35\\
Ha4 & 1.26\arcsec& 24.8&64 & 45 & 34\\
Ha8 & 1.12\arcsec& 24.8&80 & 56 & 33\\
Ha12 & 1.27\arcsec& 24.7&38 & 22 & 9\\
Ha16 & 0.99\arcsec& 24.1&31 & 23 & 8\\
ext & &&38 & 27 & 12\\
\enddata
\tablenotetext{a}{5 $\sigma$ (2.4\arcsec diameter aperture) limiting magnitudes (Vega)}
\end{deluxetable}

\clearpage
\begin{deluxetable}{cccc}
\tabletypesize{\scriptsize}
\tablecaption{Numbers of targeted and confirmed Ly$\alpha$-emitters
at $z \approx 4.5$ in each mask}
\tablewidth{0pt}
\tablehead{
\colhead{mask} & \colhead{exp time (ks)} & \colhead{targeted} & \colhead{confirmed}
}
\startdata
1 & 11.3  & 52 & 35\\
2 &  8.9 & 58 & 33\\
3 &  10.8 & 44 & 21\\
4 &  9.9 & 58 & 26\\
6 &  9.0 & 46 & 17\\
\enddata
\end{deluxetable}

\clearpage
\begin{deluxetable}{lllll}
\tabletypesize{\scriptsize}
\tablecaption{IMACS Spectroscopically confirmed Ly$\alpha$-emitters
at $z \approx 4.5$ in the LALA Cetus field\label{tbl}}
\tablewidth{0pt}
\tablehead{
\colhead{Target} & \colhead{$z^a$} & \colhead{Ly$\alpha$ flux$^b$} & \colhead{EW$_{rest}^c$(\AA)} 
}
\startdata
J020623.1-044851 & 4.431 & 2.584$\pm$0.357 &$>$105 \\
J020616.5-045457 & 4.378 & 3.795$\pm$0.410 &124$^{+481}_{-58}$ \\
J020616.2-050315 & 4.428 & 2.399$\pm$0.435 &70$^{+221}_{-33}$ \\
J020615.9-050555 & 4.380 & 2.389$\pm$0.353 &$>$82 \\
J020614.4-050523 & 4.380 & 7.448$\pm$0.376 &275$^{+1287}_{-126}$ \\
J020613.1-045101 & 4.372 & 2.648$\pm$0.401 &$>$83 \\
J020612.4-045643 & 4.385 & 3.029$\pm$0.361 &$>$118 \\
J020556.7-045854 & 4.421 & 2.850$\pm$0.359 &107$^{+477}_{-52}$ \\
J020556.5-050917 & 4.388 & 2.817$\pm$0.408 &30$^{+44}_{-8}$ \\
J020544.7-045921 & 4.382 & 2.117$\pm$0.380 &$>$78 \\
J020542.1-050520 & 4.382 & 1.993$\pm$0.377 &$>$77 \\
J020536.3-044822 & 4.414 & 5.660$\pm$0.399 &$>$179 \\
J020533.0-050042 & 4.395 & 2.844$\pm$0.408 &$>$125 \\
J020526.2-050724 & 4.390 & 2.576$\pm$0.377 &$>$67 \\
J020525.2-050048 & 4.400 & 4.276$\pm$0.431 &186$^{+1599}_{-93}$ \\
J020524.5-045241 & 4.390 & 3.221$\pm$0.420 &$>$135 \\
J020518.1-045615 & 4.406 & 4.074$\pm$0.417 &$>$161 \\
J020513.9-044638 & 4.395 & 3.268$\pm$0.431 &$>$87 \\
J020510.5-044456 & 4.406 & 3.513$\pm$0.434 &$>$70 \\
J020509.7-044005 & 4.410 & 5.030$\pm$0.372 &134$^{+292}_{-50}$ \\
J020500.1-050731 & 4.388 & 7.290$\pm$0.413 &83$^{+113}_{-18}$ \\
J020457.3-051157 & 4.398 & 5.928$\pm$0.414 &$>$116 \\
J020451.6-045737 & 4.365 & 2.221$\pm$0.386 &$>$85 \\
J020450.6-050107 & 4.378 & 2.256$\pm$0.399 &$>$87 \\
J020450.9-044323 & 4.401 & 2.124$\pm$0.414 &49$^{+107}_{-20}$ \\
J020446.9-050116 & 4.354 & 2.208$\pm$0.387 &$>$92 \\
J020438.6-044437 & 4.398 & 3.233$\pm$0.428 &43$^{+67}_{-12}$ \\
J020434.6-051016 & 4.378 & 2.813$\pm$0.410 &99$^{+411}_{-48}$ \\
J020434.5-050516 & 4.401 & 6.029$\pm$0.452 &202$^{+740}_{-90}$ \\
J020432.3-045519 & 4.360 & 2.740$\pm$0.413 &$>$97 \\
J020428.5-045924 & 4.391 & 2.766$\pm$0.393 &96$^{+338}_{-45}$ \\
J020427.4-050045 & 4.390 & 2.486$\pm$0.411 &$>$100 \\
J020425.7-045810 & 4.383 & 2.788$\pm$0.400 &$>$68 \\
J020414.3-051201 & 4.401 & 4.139$\pm$0.453 &$>$119 \\
J020413.8-044703 & 4.385 & 2.928$\pm$0.422 &$>$72 \\
J020628.0-045307 & 4.462 & 2.011$\pm$0.376 &$>$69 \\
J020623.8-044221 & 4.444 & 3.859$\pm$0.366 &$>$169 \\
J020603.5-050019 & 4.441 & 2.297$\pm$0.357 &$>$59 \\
J020601.7-050258 & 4.433 & 3.134$\pm$0.373 &$>$126 \\
J020554.3-044147 & 4.428 & 2.761$\pm$0.346 &$>$77 \\
J020550.6-045005 & 4.429 & 2.005$\pm$0.350 &69$^{+282}_{-35}$ \\
J020536.6-045520 & 4.418 & 2.689$\pm$0.393 &99$^{+632}_{-51}$ \\
J020531.4-044053 & 4.459 & 2.115$\pm$0.360 &53$^{+130}_{-23}$ \\
J020527.7-043944 & 4.436 & 3.848$\pm$0.374 &54$^{+86}_{-16}$ \\
J020519.2-044902 & 4.421 & 2.531$\pm$0.380 &$>$95 \\
J020517.5-044347 & 4.436 & 2.420$\pm$0.396 &$>$99 \\
J020513.8-043833 & 4.419 & 2.723$\pm$0.376 &$>$117 \\
J020507.2-050636 & 4.465 & 2.861$\pm$0.362 &$>$119 \\
J020506.8-043858 & 4.429 & 3.873$\pm$0.355 &87$^{+169}_{-31}$ \\
J020458.0-050140 & 4.452 & 2.155$\pm$0.377 &82$^{+411}_{-42}$ \\
J020457.3-043847 & 4.431 & 2.086$\pm$0.366 &$>$47 \\
J020452.1-043802 & 4.431 & 3.346$\pm$0.361 &$>$88 \\
J020450.2-045828 & 4.423 & 2.510$\pm$0.366 &$>$94 \\
J020448.0-044257 & 4.426 & 2.871$\pm$0.374 &79$^{+200}_{-33}$ \\
J020445.5-050008 & 4.416 & 2.297$\pm$0.383 &57$^{+139}_{-25}$ \\
J020439.0-045233 & 4.421 & 2.387$\pm$0.398 &$>$79 \\
J020438.9-044401 & 4.414 & 2.151$\pm$0.376 &$>$54 \\
J020438.9-051116 & 4.447 & 9.920$\pm$0.366 &$>$247 \\
J020434.9-045703 & 4.433 & 2.422$\pm$0.363 &48$^{+91}_{-18}$ \\
J020432.3-050916 & 4.456 & 2.992$\pm$0.373 &$>$62 \\
J020429.1-044307 & 4.434 & 3.992$\pm$0.389 &86$^{+183}_{-33}$ \\
J020424.7-043912 & 4.433 & 3.839$\pm$0.374 &$>$110 \\
J020418.1-050748 & 4.449 & 2.813$\pm$0.354 &$>$105 \\
J020617.0-045438 & 4.475 & 2.081$\pm$0.366 &$>$65 \\
J020614.0-050031 & 4.479 & 3.085$\pm$0.354 &$>$131 \\
J020612.3-045608 & 4.467 & 2.529$\pm$0.368 &$>$103 \\
J020611.1-044601 & 4.488 & 2.293$\pm$0.349 &$>$84 \\
J020603.8-044905 & 4.467 & 3.462$\pm$0.348 &47$^{+70}_{-13}$ \\
J020557.7-051003 & 4.457 & 1.901$\pm$0.352 &$>$48 \\
J020557.5-045107 & 4.461 & 2.229$\pm$0.334 &$>$85 \\
J020547.0-045744 & 4.456 & 2.785$\pm$0.357 &$>$106 \\
J020536.4-050751 & 4.484 & 4.419$\pm$0.339 &97$^{+192}_{-35}$ \\
J020535.6-044430 & 4.498 & 3.408$\pm$0.413 &$>$142 \\
J020534.6-044322 & 4.507 & 6.244$\pm$0.391 &49$^{+61}_{-9}$ \\
J020531.2-044704 & 4.459 & 2.292$\pm$0.347 &$>$95 \\
J020526.5-050310 & 4.469 & 2.377$\pm$0.346 &$>$82 \\
J020525.7-045927 & 4.487 & 1.962$\pm$0.350 &$>$53 \\
J020525.1-044552 & 4.456 & 2.688$\pm$0.351 &$>$91 \\
J020525.1-044515 & 4.444 & 1.778$\pm$0.353 &$>$51 \\
J020456.6-050145 & 4.444 & 1.796$\pm$0.351 &46$^{+113}_{-20}$ \\
J020454.8-045725 & 4.461 & 2.146$\pm$0.348 &$>$47 \\
J020451.3-045147 & 4.475 & 2.000$\pm$0.342 &$>$44 \\
J020448.3-044628 & 4.474 & 1.924$\pm$0.379 &$>$75 \\
J020442.1-050652 & 4.482 & 2.103$\pm$0.360 &$>$65 \\
J020432.5-044000 & 4.470 & 2.153$\pm$0.362 &70$^{+275}_{-35}$ \\
J020429.7-050251 & 4.462 & 2.152$\pm$0.377 &$>$70 \\
J020425.4-045610 & 4.461 & 3.000$\pm$0.361 &39$^{+60}_{-11}$ \\
J020414.6-045956 & 4.477 & 2.777$\pm$0.352 &$>$104 \\
J020618.2-045024 & 4.523 & 3.945$\pm$0.406 &55$^{+84}_{-15}$ \\
J020608.0-044623 & 4.512 & 2.374$\pm$0.414 &$>$52 \\
J020604.1-050752 & 4.495 & 2.007$\pm$0.389 &62$^{+194}_{-29}$ \\
J020557.6-043614 & 4.493 & 3.290$\pm$0.605 &$>$81 \\
J020617.0-044833 & 4.541 & 3.583$\pm$0.679 &99$^{+234}_{-40}$ \\
J020614.0-043652 & 4.561 & 4.533$\pm$0.796 &98$^{+253}_{-41}$ \\
J020603.5-045637 & 4.536 & 4.303$\pm$0.659 &27$^{+34}_{-5}$ \\
J020602.0-045633 & 4.559 & 4.398$\pm$0.661 &136$^{+409}_{-59}$ \\
J020544.1-045049 & 4.543 & 4.404$\pm$0.660 &$>$200 \\
J020503.1-044212 & 4.515 & 3.108$\pm$0.657 &136$^{+1123}_{-70}$ \\
J020534.7-051205 & 4.423 & 2.673$\pm$0.408 &$>$84 \\
J020501.7-045639 & 4.395 & 1.729$\pm$0.378 &22$^{+37}_{-8}$ \\
J020442.7-044200 & 4.408 & 1.627$\pm$0.408 &45$^{+114}_{-20}$ \\
J020441.3-044515 & 4.424 & 1.678$\pm$0.422 &$>$58 \\
J020427.1-044736 & 4.391 & 2.028$\pm$0.404 &$>$77 \\
J020414.4-044411 & 4.414 & 1.845$\pm$0.425 &$>$55 \\
J020413.5-044040 & 4.431 & 2.049$\pm$0.376 &94$^{+1938}_{-53}$ \\
J020618.2-043851 & 4.442 & 1.544$\pm$0.352 &32$^{+69}_{-14}$ \\
J020612.3-045347 & 4.442 & 1.544$\pm$0.378 &$>$52 \\
J020512.9-045025 & 4.439 & 1.890$\pm$0.383 &$>$66 \\
J020513.2-045331 & 4.484 & 2.203$\pm$0.420 &$>$63 \\
J020606.9-045021 & 4.539 & 1.364$\pm$0.383 &$>$39 \\
\enddata
\tablenotetext{a}{The redshift was measured from the wavelength of the peak pixel in the line profile from 1-d spectra. Due to the uncertainties in the wavelength calibration (see text for details), we expect the error in the measurement to be $\delta z$ $\sim$ 0.005.}
\tablenotetext{b}{The line flux was measured based on narrowband and broadband photometry, in unit of 10$^{-17}$ erg.cm$^{-2}$.s$^{-1}$.}
\tablenotetext{c}{The EW were obtained based on broadband and narrowband photometry. See \S3.3 for details.}
\end{deluxetable}

\clearpage

\clearpage

\clearpage

\clearpage

\clearpage

\clearpage

\clearpage

\clearpage

\clearpage
\clearpage

\end{document}